\documentclass[aps,showpacs,superscriptaddress,twocolumn]{revtex4}
\usepackage{amsmath,amssymb}
\usepackage{graphics,graphicx}
\usepackage{dcolumn,bm}

\newcommand{\cu}
{\affiliation{Department of Physics, University of Calcutta,
92 Acharya Prafulla Chandra Road, Kolkata 700009, India.}}

\begin{document}

\title
{ { $A+A  \rightarrow \emptyset $ model  with a bias towards nearest neighbor}}

\author{Parongama Sen}
\email[Email: ]{psphy@caluniv.ac.in}
\cu

\author{Purusattam Ray}
\email[Email:]{ray@imsc.res.in}
\affiliation{ 
 The Institute of Mathematical Sciences, CIT Campus, Taramani, Chennai 600113, India}

\begin{abstract}

We have studied $A+A \rightarrow \emptyset$ reaction-diffusion model on a ring, with a bias  
$\epsilon$ $(0 \leq \epsilon \leq 0.5)$ of the random walkers $A$ to hop towards their nearest 
neighbor. Though the bias is local in space and time, we show that it alters the universality class 
of the problem. The $z$ exponent, which describes the growth of average spacings between the 
walkers with time, changes from the value 2 at $\epsilon=0$ to the mean-field value of unity for 
any non-zero $\epsilon$. We study the problem analytically using independent interval approximation 
and compare the scaling results with that obtained from simulation. 
The distribution  
$P(k,t)$ of the spacing $k$ between two walkers  (per site)
is given by $t^{-2/z} f(k/t^{1/z})$ and is obtained both analytically and numerically.
We also obtain the result that $\epsilon t$ becomes the new time scale for $\epsilon \neq 0$.

\end{abstract}

 \pacs{05.40.Fb, 05.40.-a, 75.40.Gb}
\maketitle

%\vspace{5.5cm
\section{Introduction}

Diffusion controlled  annihilation problems have received lots of attention over the years \cite{books,bramson,torney,
lushnikov,balding,spouge,amar,avraham1,alcaraz,schutz,krebs,racz,santos,sasaki,oliveira,brunet}.
These are non-equilibrium systems of diffusing particles, which undergo reactions such as 
pairwise annihilation. Depending on the problem, these particles may represent molecules, 
biological entities, opinions in societies or market commodities and such systems are widely 
used to describe the pattern-formation phenomena in wide varieties of biological, chemical 
and physical systems. In the lattice version of the simple single species problem, each lattice 
site is filled with a particle  at time $t=0$. At each time step, the particles are allowed to jump 
to a nearest neighbor site. In general no preferred direction for the jump is assigned.  Particles 
react only when a certain number $k$ of them meet: $kA \rightarrow lA$ with $l<k$. 
Annihilating random walkers with $k=2$ and $l=0$ mimic the dynamics of voter models and 
the Glauber-Ising model in one dimension. Such systems have been studied in one dimension 
\cite{spouge,amar,avraham1,alcaraz, schutz,krebs,racz,santos}
as well as in higher dimensions \cite{kang,zumofen,peliti,droz}. The steady state of the process is 
rather simple. Depending on the initial condition, whether one starts with even or odd number 
of particles, the steady state will contain no particles or one particle respectively. The focus in all 
these analysis is how the system approaches the steady state. In particular, one intends to know 
how the number of particles decay with time and the distribution of the intervals between the 
particles evolves with time.

The dynamics of the system is governed by two processes: reaction (annihilation) and diffusion.
If the reaction time much exceeds the diffusion time, the process is reaction-limited. In this 
regime, the kinematics is dominated by the diffusion and it is well described by the mean-field 
equations. On the other hand, in the diffusion-limited regime, where the diffusion time is much 
larger than the reaction time, the process is dominated by the fluctuations caused by the reaction and 
at low dimensions, kinetics is no longer described by mean-field rate equations.
For $A+A \rightarrow \emptyset$ model ($k=2, l=0$), the critical dimension $d_c$ is 2. For 
dimension $d > d_c$, the mean-field behavior is valid which predicts that the density 
of the particles decay with time as $1/t$ for random initial configuration of the particles. 
In the mean-field picture, the time scale is 
set by the reaction rate, which at high dimensions, is given by the average steady state flux of the particles towards any particle in the system. 

At low dimensions, for $d \leq 2$, the problem of recurrence of random walks appear.  From 
the point of diffusion, the collision rate is effectively infinite. The  rate equation gives the 
asymptotic behavior of the density of particles $N_p(t)$ decaying as $\sim 1/ {t^{1/z}} $ 
for $d < d_c$ with $z=2$. The average domain size or interval (i.e. the distance between neighboring 
walkers in one dimension)  $D_s$ scales as $t^{1/z}$ and this is the only length scale which 
characterizes the reactant distribution. The scaling is robust as long as the particle motions 
are uncorrelated, diffusive with well defined mean and fluctuation.  At $d=d_c$, the mean-field result 
is retrieved with logarithmic corrections. The value of $d_c$ and the behavior of $N_p(t)$ have 
been conjectured by scaling arguments \cite{kang}, exact results in one dimension 
\cite{amar,avraham1,racz,privman}, 
probabilistic approaches \cite{spouge,balding} and renormalization group calculations 
\cite{peliti,droz,ohtsuti,lee}.

Here, we present the study of the time evolution of a set of randomly distributed random walkers on a ring,  
having the interaction $A + A \to \emptyset$, evolving with the following dynamical rule: at each
time step each walker moves towards its nearer neighbor with a probability $1/2 + \epsilon$. 
$\epsilon = 0$ 
would give the usual unbiased random walkers while for $\epsilon = 1/2$ the walkers will always 
move towards their nearer neighbors making the system quasi-deterministic. 
When the two  neighbors are at the same distance the particle 
moves either way with equal probability. The ring geometry is taken to 
impose periodic boundary condition. We call this model 
the anisotropic  walker model (AWM) hereafter. The AWM is motivated by the 
 social phenomenon of opinion formation and  for $\epsilon = 1/2$, coincides  with 
the binary opinion dynamics/spin model (BS model)  proposed in \cite{biswas-sen}. 

The  BS model was proposed to mimic opinion formation in a  society where 
the opinions are binary.  Here 
 an agent's opinion is decided   by the size of the  neighbouring domains
(in a domain  all opinions are of the same type) which may be interpreted as social pressure.  
In the BS model, surprisingly, it was found that $z \simeq 1$. That means $\epsilon$ alters the 
universality class of the problem. The generalization of the model
with $\epsilon > 0$ implies that an agent in the BS model follows the 
opinion of the larger domain with a probability (larger than 0.5) which makes the 
system fully stochastic. 

Simulations of  AWM  model  with rather small sizes indicated \cite{BSR} that any $\epsilon$ in 
the range $0 < \epsilon \le 1/2$  alters the value of the exponent $z$ 
compared to the case $\epsilon = 0$ 
\cite{footnote}. 
Here, we study AWM to understand the effect of $\epsilon$ on the long time behavior of the 
$A+A \rightarrow \emptyset$ model. We study, particularly, the distribution $P(k,t)$ 
of the interval sizes (the distance between the neighboring walkers equivalent to the domains in the opinion formation model) $k$ per site at time 
$t$ and its evolution with time. This distribution is of importance 
as it helps analyzing the dynamic process  and has been calculated in many dynamical  models in one dimension.
Often the mapping with Glauber  spin picture is utilized to 
evaluate this function. The present model  however, is  not 
equivalent to a Glauber like model and thus one may  expect the results to be  
different in general. 

We obtain the scaling solution of $P(k,t)$ for late times; it is of interest to check whether nonzero values of $\epsilon$ 
can alter the known form for $\epsilon=0$. We have employed Independent Interval Approximation (IIA) 
(to be described below) to 
study the evolution analytically and complemented the findings with Monte Carlo simulation results.

\section {IIA Analysis}

% \cite{abraham,alemany},
%simple diffusion \cite{satya} and zero temperature Glauber dynamics for Ising, Potts model \cite{derrida1,derrida2} and several reaction diffusion systems \cite{krapivsky}.
The independent interval approximation (IIA) was originally proposed in \cite{old} and has been applied to  several
studies of the diffusion limited processes.  
IIA implies that the intervals or gaps between the nearest neighboring particles are 
independent of each other.
 IIA has been successfully applied to the case of diffusion 
limited  annihilation $A + A \to \emptyset$ (which maps to the Glauber spin model) to obtain the 
inter-particle interval distribution function \cite{avraham2}.
%Here, the exact expression of the equal-time two spin correlation function has
%also been used to derive the result.
In  simple diffusion problems,  the idea of IIA has been used to predict the persistence 
exponents in excellent agreement with the simulation results \cite{satya,dornics}.

The assumption  that the intervals or domains are uncorrelated 
has been later  developed self-consistently for several models in \cite{krapivsky}.
In particular, the IIA analysis describes the dynamics of the model for $\epsilon=0$ extremely well  \cite{avraham2,derrida,krapivsky} qualitatively.
The quantitative accuracy increases if one  uses  the exact expression of the equal-time two spin correlation function alongwith IIA \cite{avraham2}.
We show here that the analysis can be extended to non-zero $\epsilon$ case also. IIA analysis gives the scaling form for $P(k,t)$ with the scaling exponent $z=1$ and also gives the exponential decaying
form of the scaling function. IIA results are well supported by numerical simulation results. 
 Our results show that $\epsilon t$ becomes the new time-scale
for non-zero $\epsilon$. 
As a result, at low $\epsilon$, it takes longer time to reach the asymptotic 
scaling limit.

Within IIA, the master equation that describes the rate of change of $P(k,t)$ (written $P(k)$ for brevity) with 
time can be broken into $\epsilon$-independent and $\epsilon$-dependent terms and 
is given by 

%\begin{small}
\begin{equation}
\frac{dP(k)}{dt}  =  I_1(k) + 2 \epsilon I_2(k) 
\label{master}
\end{equation}
where,
\begin{eqnarray}
%I_1  & = & P(k+1) + P(k-1) - 2P(k) - \frac{P(k)P(1)}{\sum_k P(k)}  - \frac{P(k-1)P(1)}{\sum_k P(k)}  \nonumber \\
%& & - \frac{P(1)}{(\sum_k P(k))^2} \bigg[ P(1)P(k-2) + \sum_{m > 1}^{k-2} P(m) P(k-m-1) \bigg] \nonumber 
I_1(k) & = & P(k+1) + P(k-1) - 2P(k) + \frac{P(1)}{N^2} \nonumber \\ 
& & \bigg[ \sum_{m=1}^{k-2} P(m)P(k-m-1) - N ((P(k)+P(k-1)) \bigg] \nonumber \\
& & {} {} 
%\nonumber
\label{master_1}
\end{eqnarray}

\noindent and

\begin{eqnarray}
I_2(k) & = & \frac{P(k+1)}{N}\bigg[ \sum_{m > k+1} P(m) - \sum_{m < k+1} P(m) \bigg] \nonumber \\
& & + \frac{P(k-1)}{N} \bigg[ \sum_{m < k-1} P(m) - \sum_{m > k-1} P(m) \nonumber \\
& & - P(1) \bigg](1-\delta_{k,2}) - \frac{P(k)P(1)}{N^2} \sum_{m > 1} P(m) + \frac{P(1)}{N^2} \nonumber \\
& &  \bigg[ \sum_{m > 1}^{k-2} P(m)P(k-m-1) - P(1)P(k-m-1) \bigg] \nonumber \\
& & - P(1)\bigg[1 - \frac{P(1)}{N} \bigg] \delta_{k,2} \nonumber \\
& & {} {} 
%\nonumber 
\label{master1}
\end{eqnarray}  

%\end{small}

\noindent where $N = \sum_k P(k,t)$, is the density  of the intervals (number of intervals 
 per lattice site) at time $t$ and is equal to the density of the particles $A$ at time $t$. 
Naturally, $\sum_k k P(k) = 1$, which comes from the conservation of the total length of all the intervals.
The details of the derivation of the equation (\ref{master}) are given in the Appendix.

The $\epsilon = 0$ case, which corresponds to $I_1(k)$ in eq. (\ref{master}) has been studied 
before using IIA \cite{krapivsky}. $P(k,t)$ is found to have the expected scaling form: 
\begin{equation}
P(k,t) = t ^{-2/z} f(\frac{k}{t^{1/z}}). 
\label{scaling1}
\end{equation}
with the scaling exponent $z=2$. The  scaling function $f(x) \sim \exp(-\alpha x)$ at large $x$. 
The average length $k$ of the intervals at time $t$:
$\langle k(t) \rangle = \int k I_1(k)dk/\int I_1(k) dk \propto t^{1/z}$.  The scaling behavior 
given by the eq. (\ref{scaling1}) matches 
with that obtained from the exact analysis of the model \cite{derrida} except for the value of $\alpha$. 
It is to be noted that 
%$\sum_k I_1(k) = N$, $\sum_k k I_1(k) = 1$ (normalization condition) and 
$ \frac{dN}{dt} = -2P(1)$. The last result implies that any change in 
$N$ is brought out by the annihilation of two $A$ particles which were at a unit distance apart and coalescence of the intervals separated by these two particles.

\begin{figure}
\begin{tabular}{cc}
\includegraphics[width= 0.6\columnwidth,angle=270]{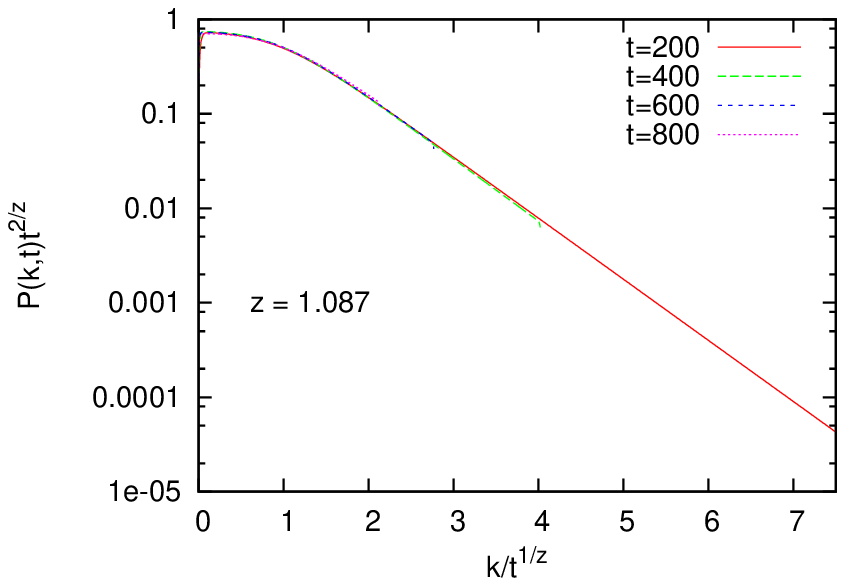}\\
\includegraphics[width= 0.6\columnwidth,angle=270]{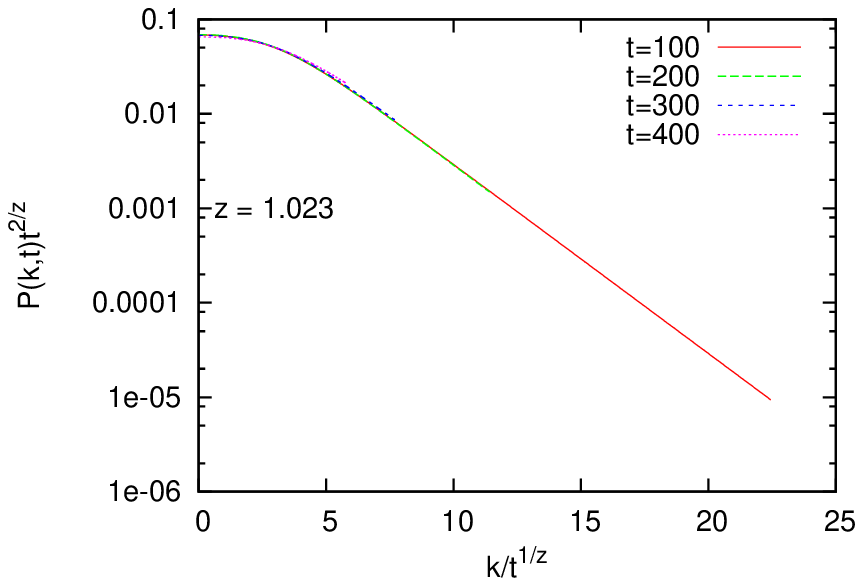}
\end{tabular}
\caption{(Color online) The scalings of the interval size distributions $P(k,t)$ obtained from IIA 
for $\epsilon = 0.1$ and 0.5 are shown at different times $t$}.
\label{fig:scaling1}
\end{figure}

For nonzero $\epsilon$, the term $I_2(k)$ appears in the rate equation. It is to be noted that now 
%$\sum_k I_2(k) = 0$, $\sum_k k I_2(k) = 0$ and 
%$\sum_k \frac{dI_2(k)}{dt} = -\epsilon P(1)[1+\frac{P(1)}{N}]$. 
$ \frac{dN}{dt} = -2P(1) -\epsilon P(1)[1+\frac{P(1)}{N}]$. 
 We solve eq.~({\ref{master}) numerically 
%using Euler's method. We 
starting  with intervals of  
sizes 1,2 ...$n$ with distribution 
$P(k,0) = \eta_1 \delta_{k,1}  + \eta_2 \delta_{k,2} + ...+ \eta_n \delta_{k,n}$, where 
$\eta_1, \eta_2 ...\eta_n$ are random numbers between 0 and 1 and  
$\eta_1 + 2\eta_2 +...+ n\eta_n=1$. 
We find that the final result is insensitive to the choices of $\eta$'s or the number of different 
types of intervals to start with or different configurations of the starting interval distribution. 
Most of our analytical results are obtained with initial intervals of size $n \leq 3$ and 
 averaged typically over 10 different initial configurations. On the other hand, we find 
that the results depend crucially on the discrete time step involved in Euler's method 
and the observation time. 
In most of our results, time is incremented by $\delta t = 0.01$ at each step. 
%The distributions are generated at subsequent times using Euler 
%method. 
We have studied systems of sizes $L = 1000$ and 2000. We find that this gives us a good idea of the 
validity of the scaling and the exponential decay of the scaling function at large arguments at the expense 
of a reasonable computer time. It may also be added that the value of $L$ enters
the numerical calculation indirectly as the rate equation is in terms of 
probabilities and $L$ determines only the  upper bound of the 
size of the domains. 
%We have found no appreciable finite size effect on the results.  
We check that $z$ approaches the values 2 and 1 at $\epsilon = 0$ and 0.5 respectively 
as $\delta t $ is lowered for $L=2000$ and there is no appreciable finite size effect.

\begin{figure}
\includegraphics[width= 0.6\columnwidth, angle=270]{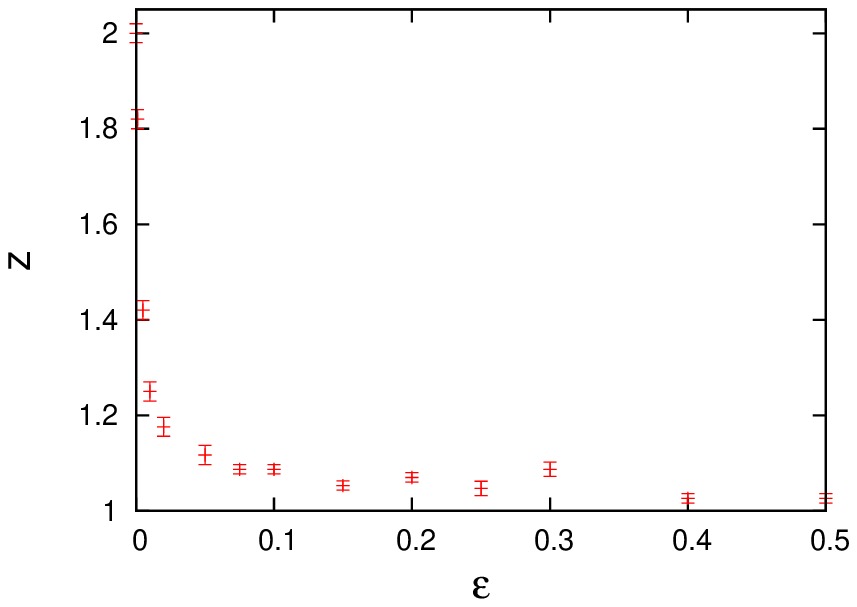}
\caption{(Color online) Values of $z$ (along with error bars) obtained using IIA is plotted against $\epsilon$.  }
\label{zfromIIA_2}
\end{figure}

For nonzero $\epsilon$, solution of eq.~(\ref{master}) obtained numerically shows that $P(k,t)$ retains the 
same scaling form as in eq.~({\ref{scaling1}). 
One can obtain a collapse by suitably scaling the variables using trial values of $z$ for each $\epsilon$. 
Fig.~\ref{fig:scaling1} shows the scaling for 
two specific values of $\epsilon $ at four different times. 
%With $\delta t = 0.01$, the value 
%of the scaling exponent $z$ is overestimated even for $\epsilon = 0$ for which the known value of $z$ is equal to 2. 
% and 0.5.  
The dependence of $z$ on $\epsilon$ is shown in Fig.~\ref{zfromIIA_2}. 
For $\epsilon < 0.1$, $z$ shows a relatively sharp increase to $\sim 2.0$ as $\epsilon \to 0.0$. 
However, above $\epsilon = 0.1$, the variation is not systematic 
which suggests that the value is actually a constant. The values of $z$
in this region differ from 1 by not more than ten percent. The data collapse using using 
eq.~({\ref{scaling1}) which we have used to extract the value of $z$ is insensitive to this small fluctuation 
in the value. In principle, one can think of a functional dependence of $z$ on $\epsilon$ but other
results (such as the scaling behavior  in eq.~({\ref{qkequation}) appearing later in the paper)  indicate that the  variation of $z$ for $\epsilon >0$ is only an artefact of the sudden change 
in $z$ at $\epsilon=0$ and actually $z$ is a constant in this region.
This is supported by the fact that the  sharp rise of $z$ occurs at  lower $\epsilon$ values as one 
increases the observation time (in the scaling regime). 
%We show in Fig. \ref{zfromIIA_1} the typical scaling behaviour for small and large observation times for $\epsilon = 0.2$ indicating  that the exponent decreases  
%at larger times. The collapse of the different curves is also of much better quality at large observation times. 
We thus conclude that the IIA method gives $z=1$ for all $\epsilon \neq 0$ 
and 2 only when $\epsilon = 0$,  consistent with earlier results 
obtained from simulations in \cite{BSR}. We also show below that  
 for $\epsilon > 0$, the cumulative distribution gives consistent results 
with that obtained theoretically, using $z=1$, 
 to support our conclusion.  
%%as one finds from IIA calculation.  

\begin{figure}
\noindent \includegraphics[clip,width= 6cm, angle=270]{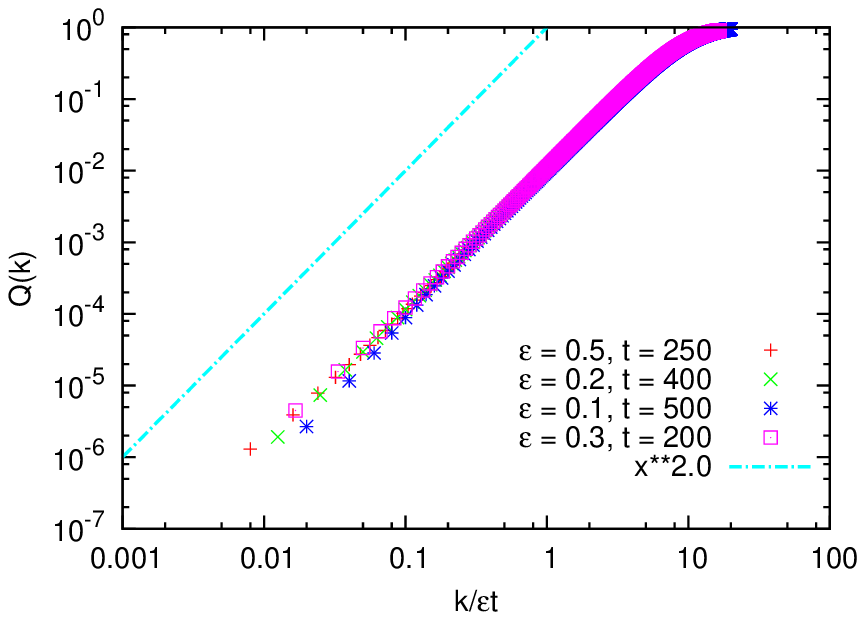}
\caption{(Color online) The cumulative distribution $Q(k,t)  = \int{_0}^k nP(n)dn $ from IIA for nonzero
$\epsilon$. The straight line has slope = 2.0.}
\label{cumuIIA}
\end{figure}

%\begin{figure}
%\begin{tabular}{cc}
%\includegraphics[width= 0.5\columnwidth,angle=270]{scaling002_first.eps}\\
%\includegraphics[width= 0.5\columnwidth,angle=270]{scaling002_last.eps}\\
%\end{tabular}
%\caption{The scalings of the interval size distributions $P(k,t)$ from IIA 
%for $\epsilon = 0.2$ are shown for initial and late time steps.
%The exponent $z$ decreases towards unity at larger times.}
%\label{zfromIIA_1}
%\end{figure}

Using  the form  of $P(k,t)$ given by eq.~(\ref{scaling1}), 
one can calculate the
cumulative distribution $Q(k,t) = \sum_{m=1}^k mP(m,t)$. Assuming an 
exponential behaviour of the scaling function ($f(x) \sim \exp(-\alpha x)$), one gets
$Q(k,t) =  1-\exp(-\alpha k/t^{1/z})(1+ \alpha k/t^{1/z})$.
In Fig.~\ref{cumuIIA} we plot the cumulative distribution $Q(k,t) = \sum_{m=1}^k mP(m)$ as 
obtained from IIA calculations for different 
values of $\epsilon$ and times $t$. The distribution  exhibits the scaling
$ Q(k,t) \sim h(\frac{k}{t})$ which is consistent with the value $z=1$. 
The scaling function appearing in the cumulative distribution 
behaves like $h(x) \sim x^{\delta}$ for small $x$ and goes to unity at large 
$x$. The value of $\delta$ is close to 2 which  agrees with the theoretical estimate  when $k/t^{1/z}$ is small.  
%as we have noted earlier that the scaling function $f$ in eq. (\ref{scaling1}) has an exponential decay.
We note another interesting feature; the curves for different values of $\epsilon$
collapse when the data are plotted against $k/(\epsilon t)$, such that the 
behavior of $Q(k,t)$ may be written as 
\begin{equation}
 Q(k,t) \sim h(\frac{k}{\epsilon t}). 
\label{qkequation}
\end{equation}
The scaling shows that for any non-zero $\epsilon$, $k \sim t$, that is $z=1$.
We will get back to this behaviour later in section \ref{disc}.

\section {Simulation Results}

We verify the scaling results by Monte Carlo simulation. We start (at $t=0$) with a one dimensional 
chain of size $L$ with half of the lattice sites occupied randomly by the particles $A$. 
$L$ is varied between $10^4$ and $10^5$ and periodic boundary condition is used. 
In a single update, we choose a site randomly and if there is a particle
its  position is  
updated. $L$ such updates constitute  one Monte Carlo step and the  
 dynamics is  asynchronous. 
The distances (in terms of lattice units) of 
the neighboring particles are obtained and the particle is shifted one lattice site left or right 
with probability $(1/2+\epsilon)$ towards the nearer neighbor and $(1/2-\epsilon)$ along the 
other direction. If the new site is occupied, then both the particles occupying that site and the 
one which has hopped over to it are removed from the system. 
 As $\epsilon$ is made larger, 
the rate of annihilation become high and as a result 
very few walkers remain at large times. It poses difficulty in obtaining  good statistics of the data 
for the distribution $P(k,t)$. One has to carefully identify the scaling regime which is 
almost nonexistent for small systems. Hence, for this analysis, $L =  5\times 10^5$ was used 
for which the scaling regime can be obtained only for small values of time $t$.  
For small $\epsilon$-values, a system size of $L = 10000$ suffices. 
The data is averaged typically over 1000 different random initial configurations of the positions 
of the particles.
%The number of configurations is varied accordingly   (10 for $\epsilon = 0.5$ and 
%1000 for $\epsilon = 0$). 
For $\epsilon =0$, $P(k,t)$ follows eq.~(\ref{scaling1}) as in IIA with  $z = 2.0$. 

Before discussing  $P(k,t)$ for nonzero $\epsilon$, we check that for the large system sizes considered, 
the fraction of surviving walkers 
shows the scaling $N_p \propto  t^{-1}$ (see Fig. \ref{dynamics}) and there is no need to consider any correction to scaling (for any value of $\epsilon$) reported earlier \cite{BSR}  for comparatively  smaller sizes.  
The results for $P(k,t)$ are plotted in Fig. \ref{fig:scaling2}.
%%%%%%%%%%%%%%%%%%%%%%%%%%%%%
\begin{figure}
%\noindent \includegraphics[clip,width= 3.5cm, angle=270]{Ann_dir.eps}
\noindent \includegraphics[clip,width= 5cm,angle=270]{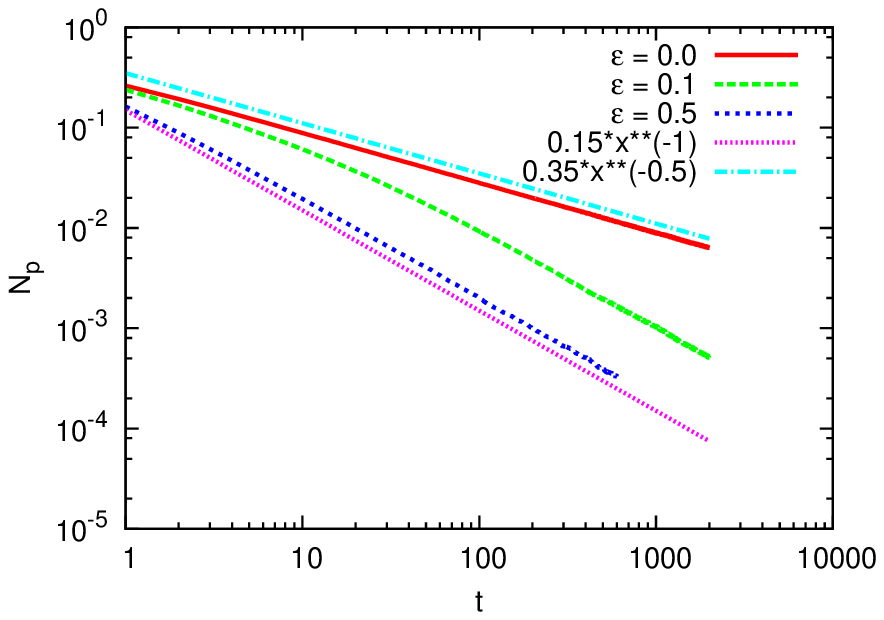}
\caption{
(Color online) $N_p$ versus $t$ from simulations for different values of $\epsilon$.}
\label{dynamics}
\end{figure}
%%%%%%%%%%%%%%%%%%%%%%%%%%%%%
%%%%%%%%%%%%%%%%%%%%%%%%%%%%%
\begin{figure}
\begin{tabular}{cc}
\includegraphics[width= 0.6\columnwidth,angle=270]{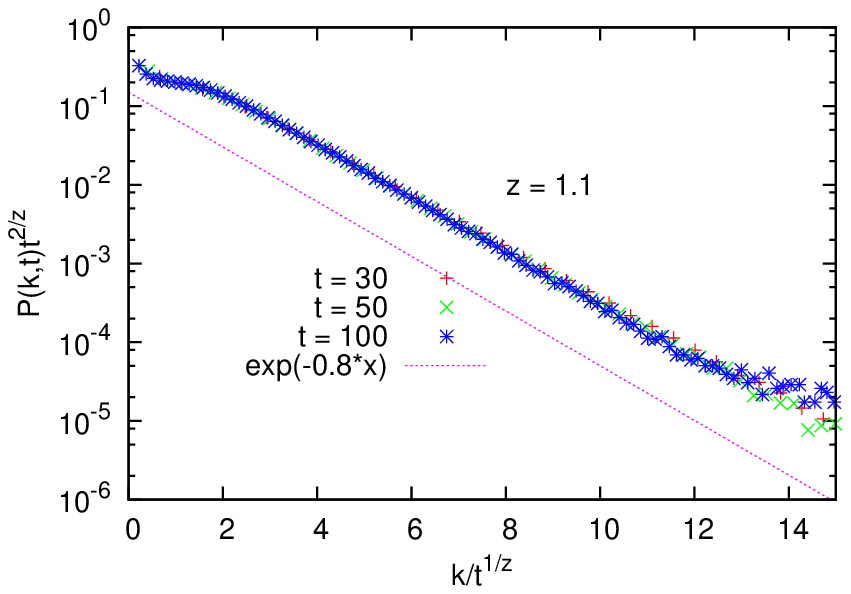}\\
\includegraphics[width= 0.6\columnwidth,angle=270]{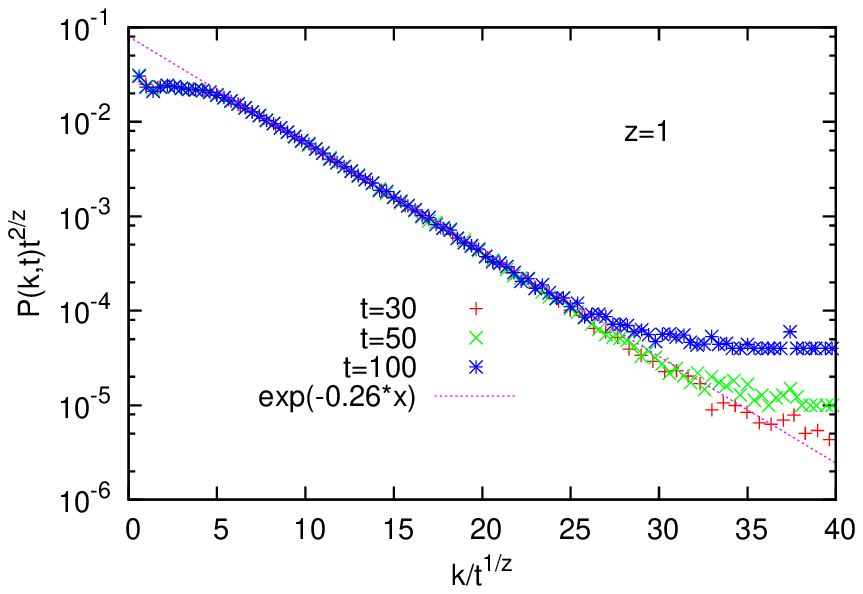}
\end{tabular}
\caption{(Color online) The scalings of the interval size distributions $P(k,t)$ from simulation
for $\epsilon = 0.1$ and 0.5
at times $t = 30, 50$ and 100 are shown.  The scaling holds good for $1/z = 0.91$ and $1/z = 1.0$ respectively.}
\label{fig:scaling2}
\end{figure}
%%%%%%%%%%%%%%%%%%%%%%%%%%%%%
We report  results for $\epsilon \geq  0.1$
which show agreement with the scaling behavior as given in eq.(\ref{scaling1}) (shown in Fig. \ref{fig:scaling2}).
The value of $z$  shows deviation from unity only for the smallest value of $\epsilon$. 
We also estimate the cumulative function $Q(k,t)$ which again shows a  collapse when plotted against $ x = k/\epsilon t$ for 
$x > 1$ (see Fig. \ref{cumusimu}). However, although $Q(x) \propto x^\delta$ as obtained in IIA, the exponent $\delta$ has a lower
value $\sim 1.8$.
We also make a further analysis: if eq.~(\ref{qkequation}) is obeyed with 
$h(x) \propto x^\delta$, $\log[Q(x))/Q(x/b)] $ must be equal to  
$\delta \log b$ 
where $b$ is a scaling factor. 
Estimating $\delta$ in this way, we find $\delta \approx 1.78$ (see inset of Fig. \ref{fig:scaling2}). This plot has been done for small values of 
$x$ where the power law behaviour is expected to be valid; for large
values of the argument, $Q(x)$ approaches unity as it is a cumulative 
probability. 

\section{Discussions and concluding remarks}
\label{disc}

In this paper, we have used two approaches to study the AWM model. 
The results of the IIA approach and the simulations agree quite well but 
the values of $\alpha$ (associated with the scaling function $f$) differ notably (e.g., $\alpha$ for $\epsilon = 0.5$ is approximately 0.45 from IIA while
simulations give a value $\sim 0.26$). This, however, is not surprising as even for $\epsilon = 0$,
$\alpha  \approx 0.55$ \cite{krapivsky} is quite different from the exact result ($0.368468$) \cite{derrida}.

Another difference which appears is the disagreement of the value of $\delta$ (associated with the scaling function $h$)
in the two methods. The theoretical value $\delta =2$ is derived 
assuming the scaling function $f$ occurring in eq. (\ref{scaling1}) has an exponential decay  which is true for 
large values of the argument of $f$ in both cases. The discrepancy in the value of $\delta$ thus suggests that for small values of the argument there may be a significant difference in  the  form of the scaling function in the   IIA and the simulation results. However, this region where 
the difference  is speculated to occur is 
rather narrow to make a systematic study. 

The result that $k/\epsilon t$ appears as the scaling variable,  obtained in both the approaches,
immediately suggests that $\alpha$ in eq.~(\ref{scaling1}) varies as $ 1/\epsilon$. We note the 
values of $\alpha \epsilon$ to check whether this is true and find good agreement for the IIA values
for $\epsilon \geq 0.3$ and very good agreement for the values obtained in simulation
for $\epsilon > 0.1$. $\alpha \epsilon \sim 0.13$ for $\epsilon > 0.1$ (from the simulation results), while it apparently 
decreases for lower values of $\epsilon$. However, as we have noted earlier, 
the results for very small $\epsilon$ shows the effect of the $\epsilon = 0$ point which 
belongs to a different universality class. 
Since $\alpha$ value obtained from the simulation happens to be 
more reliable, we conclude that indeed $\alpha \propto 1/\epsilon $ for all values of $\epsilon > 0$.

\begin{figure}
\noindent \includegraphics[clip,width= 8cm]{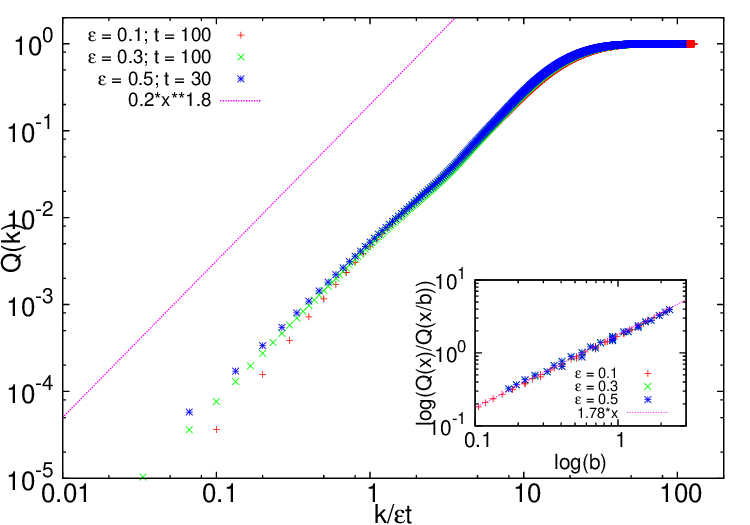}
\caption{(Color online) The cumulative distribution $Q(k) = \int{_0}^k nP(n)dn $ for $\epsilon=0.10 ;t=100$, $\epsilon=0.30;t=100$,  
and $\epsilon = 0.5; t = 30$ is plotted against  $k/{\epsilon t}$. The straight line has slope = 1.8. The inset shows the log-log plot of the ratio 
of two values of $Q$ at $x$ and $x/b$ against $b$ where $b$ is a scaling factor.}
\label{cumusimu}
\end{figure}

To summarise, we  have considered the 
$A+A \rightarrow \emptyset$ reaction-diffusion model on a ring, with a bias
$\epsilon$ $(0 \leq \epsilon \leq 0.5)$ of the random walkers $A$ to hop towards their nearest
neighbor. The interval size distribution $P(k,t)$ is evaluated using the IIA method and compared 
to results obtained from numerical simulations. Both the 
methods show that for $\epsilon \neq 0$, the exponent $z = 1$ in contrast to 
$z = 2$ for $\epsilon = 0.0$. The raw data may not give the value of 
$z$ for $\epsilon \neq 0$ very accurately in IIA, but the cumulative distribution function 
$Q(k,t)$ shows that the scaling variable is indeed $k/t$ for $\epsilon \neq 0$. 
The exponential form of the scaling function $ f \sim \exp(-\alpha x)$ for all $\epsilon$-values is also obtained by IIA calculation and in simulation. The value of $\alpha$ however does  not match 
between the IIA calculation and simulation for any $\epsilon$. Simulation shows that 
$\alpha \sim 1/\epsilon$ for $\epsilon > 0.1$. We guess that this is true for all $\epsilon$. 
Finally, we note that $\epsilon t$ enters the scaling argument implying a $1/\epsilon$ dependence of the time scale in the system.

As has been mentioned in the Introduction, the  diffusion and pairwise annihilation model has been 
studied a lot in the past in the context of modelling chemical reactions \cite{house}. 
Systems of reacting particles 
are typical of complex irreversible nonequilibrium systems. It is crucial to ask what determines the universality class of the diffusing-annihilating particle system which 
is probably the simplest interacting particle system. Our 
study directly focusses on that. We show that our AWM-model yet again gives rise to a critical dynamics 
 as the system approaches towards the steady state. We show that the exponents describing the dynamics 
changes form the value 2 to 1 as soon as the bias $\epsilon$ is introduced. 
%Although there is agreement regarding the value of $z$ in the two methods, 
%the scaling function has completely different behaviour, the IIA indicating 
%an exponential decay while the simulations indicate a power law behaviour. 
%The exponential decay is of course valid for $\epsilon = 0$ although it is known that IIA does not give quantitatively correct result for the scaling function even in this limit.  As IIA is a rather gross approximation for
%the present problem, we would  accept the power law behavior 
%found in the simulations. It  seems quite realistic that   for $\epsilon 
%\neq 0$ 
%the scaling function  has a different behaviour compared to that for $\epsilon = 0$
%as they belong to different universality classes for these two regimes. 

\noindent Acknowledgements: Inspiring discussions with 
S. Biswas, 
P. Krapisvsky, 
R. Redner  
and
P. Shukla 
are acknowledged.
 Financial support from CSIR project is acknowledged by PS. 
The authors also thank Institute of Mathematical Sciences (IMSc) associateship program.

\vskip 0.5cm

\appendix
\section{Derivation of equation (\ref{master}).}

%Derivation of equation:
We consider the probability 
$P(k, t + \Delta t)$. It will have contributions from several phenomena. 
Whenever we consider the movement of a domain wall, we have to compare the 
sizes of the two domains neighbouring it. If the sizes are equal, the 
probability of a move to either side is simply 1/2. 

The probability that a domain remains same in size is 
$(1- 2\Delta t) P(k)$. A domain of size $k+1$ may reduce to a domain of 
size $k$ if either of its two edges moves so as to shrink its size by 1. This
probability will depend on the size of the adjacent domain, say $m$. If 
$m >  k+ 1$ such a move will happen with probability $(1/2 + \epsilon)$. 
and with  probability $(1/2 - \epsilon)$ otherwise. 
%If $m = k$, such a move takes place with probability 1/2. 

A domain of size $k-1$ can also grow to a domain of size $k$. Once again, one has to take care of the size of the adjacent  domain, $m$. 
It is convenient  to consider the two cases $k > 2$ and $k =2$ separately
here. $k =1$ will obviously not have any contribution from this process.
For $k > 2$, moves will depend on whether $1< m <  k-1$ in which case the probability is $(1/2 + \epsilon )$ 
while for  $m > k-1$ the probability is  $(1/2 - \epsilon )$. 
One has to ensure that $m > 1$ in the first case as for $m=1$, a domain
annihilation will take place.
For $k = 2$, one has a domain of length unity growing to a domain of length 
2 and this will  be possible only for $m 
\neq 1$ and with probability $(1/2 - \epsilon )$.  

A loss term will occur for the case when an adjacent domain of size one gets annihilated and this occurs with probability $(1/2 + \epsilon )$ when 
another domain of size $m > 1$ is its neighbour. If $m = 1$, this occurs with probability 1/2.
A gain term will also be there when a two domains get annihilated and a domain of size $k$ results in the process.   
Using the shortened notation $\epsilon_{+} =1 +  2\epsilon$;
 $\epsilon_{-} =1 -  2\epsilon$ ;
$k_{+} = k + 1$;
$k_{-} = k -1$,
and taking care of all these terms, one gets

\begin{widetext}

\begin{eqnarray}
P(k, t + \Delta t)  & = & (1-2\Delta t)P(k)  \nonumber \\ 
  &  +  & \Delta t \frac{P(k+1, t)}{\sum_k P(k,t)}  \bigg [ \epsilon_{+}
 \sum_{m > k_{+}} P(m) + \epsilon_{-} \sum_{m < k_{+}} P(m) + P(k_{+}) \bigg]     \nonumber  \\
 &  + & \Delta t \frac{P(k_{-})}{\sum_k P(k)}  \bigg[  \epsilon_{+} \sum_{1<m < k_{-}} P(m) +  
\epsilon_{-} \sum_{m > k_{-}}  P(m) + P(k_{-}) \bigg ] 
  (1-\delta_{k,2})    \nonumber \\
 &  + & \Delta t P(1)\bigg[1-P(1)/\sum_k P(k)\bigg]   \epsilon_{-} \delta_{k,2}   \nonumber \\
&  - &   \Delta t \frac{P(k)P(1)}{(\sum_k P(k))^2}  \bigg[ \epsilon_{+} \sum_{m > 1} P(m) + P(1) \bigg ]   \nonumber  \\
&   +  & \Delta t\frac{P(1)}{(\sum_k P(k))^2}  \bigg[ P(1)P(k-2) + \epsilon_{+} \sum_{m > 1}^{k-2} P(m) P(k_{-} -m) \bigg]  \nonumber \\
%+ \nonumber\\ 
%& &2 \epsilon \sum_{m > 1}^{k-2} P(m) P(k_{-} -m) \bigg ]  \nonumber \\
%&  &   + \frac{P(k-1)P(1)}{(\sum_k P(k))^2}  \bigg [ 2(\frac{1}{2} + \epsilon) \sum_{m < k-1} P(m) + 2(\frac{1}{2} - 
%\epsilon) \sum_{m > k-1} P(m) + P(k-1) \bigg]  \nonumber \\
\end{eqnarray}
\end{widetext}
For $\epsilon = 0$, the 2nd and third term can be rewritten as a single term 
and without using the Kronecker $\delta$s.


\begin{thebibliography}{}
\bibitem{footnote} Note that in \cite{BSR} $\epsilon$ was defined slightly
differently; $\epsilon = 1.0$ corresponded to the BS model while 
$\epsilon = 0.5$ the pure random walk model.
\end{thebibliography}
\end{document}